\documentclass[aps,showkeys]{revtex4}

\usepackage{epsfig}
\usepackage{eepic}
\usepackage{latexsym}
\usepackage{rotating}
\usepackage{lscape}
\usepackage{graphics}
\usepackage{dcolumn}% Align table columns on decimal point
\usepackage{bm}% bold math
\newcommand{\beq}{\begin{equation}}
\newcommand{\eeq}{\end{equation}}
\newcommand{\bd}{\begin{displaymath}}
\newcommand{\ed}{\end{displaymath}}

\newcommand{\bei}{\begin{itemize}}
\newcommand{\eei}{\end{itemize}}

\newcommand{\bee}{\begin{enumerate}}
\newcommand{\eee}{\end{enumerate}}

\begin{document}

\noindent
\title{Air density dependence of the response of the PTW SourceCheck 4pi ionization chamber for $^{125}$I brachytherapy seeds}

\author{J. Torres del R\'{\i}o$^1$, A. M. Tornero-L\'opez$^2$, D. Guirado$^1$, J. P\'erez-Calatayud$^{3,4}$, A. M. Lallena$^5$ \\
{\small {\it 
$^1$Unidad de Radiof\'{\i}sica, Hospital Cl\'{\i}nico Universitario San Cecilio, E-18016 Granada, Spain.\\
$^2$Servicio de Radiof\'{\i}sica y Protecci\'on Radiol\'ogica, Complejo Hospitalario Universitario de \\Gran Canaria Dr. Negr\'{\i}n. E-35010 Las Palmas de Gran Canaria, Spain.\\
$^3$Unidad de Radiof\'{\i}sica en Radioterapia. Hospital La Fe. E-46026 Valencia, Spain.\\
$^4$Cl\'{\i}nica Benidorm. E-03501 Alicante, Spain.\\
$^5$Departamento de F\'{\i}sica At\'omica, Molecular y Nuclear, Universidad de Granada, E-18071 Granada, Spain.
}}}

\date{\today}

\bigskip

\begin{abstract}
\noindent
{\it Purpose}: To analyze the air density dependence of the response of the new SourceCheck 4pi ionization chamber, manufactured by PTW.\\
\noindent
{\it Methods}: The air density dependence of three different SourceCheck 4pi chambers was studied by measuring $^{125}$I sources. Measurements were taken by varying the pressure from 746.6 to $986.6\,$hPa in a pressure chamber. Three different HDR 1000 Plus ionization chambers were also analyzed under similar conditions. A linear and a potential-like function of the air density were fitted to experimental data and their achievement in describing them was analyzed. \\
\noindent
{\it Results}: SourceCheck 4pi chamber response showed a residual dependence on the air density once the standard pressure and temperature factor was applied. The chamber response was overestimated when the air density was below that under normal atmospheric conditions. A similar dependence was found for the HDR 1000 Plus chambers analyzed. A linear function of the air density permitted a very good description of this residual dependence, better than with a potential function. No significant variability between the different specimens of the same chamber model studied was found. \\
\noindent
{\it Conclusion}: 
The effect of overestimation observed in the chamber responses once they are corrected for the standard pressure and temperature may represent a non-negligible $\sim 4\%$ overestimation in high altitude cities as ours (700 m AMSL). This overestimation behaves linearly with the air density in all cases analyzed.
\end{abstract}


\maketitle

\section*{Introduction}

Currently, brachytherapy with radionuclides emitting low energy photons is a very common technique in which both permanent (for treating prostate, brain and breast cancers) and temporary (in case of ophthalmological and lung tumors) implants are used. For example, $^{125}$I interstitial seed-implant brachytherapy is an effective treatment for localized prostate cancer in which, typically, from 60 to 100 seeds are transperineally implanted. Before therapy is carried out, a quality control of the seeds is recommended in order to provide a safe and effective treatment that fulfills the requirements of the dosimetric prescription. 

International institutions have provided general recommendations regarding the pre-implant quality control of the seeds. In particular, those of the American Association of Physicists in Medicine (AAPM) \cite{AAPM} and the European Society for Radiotherapy and Oncology (ESTRO) \cite{ESTRO} for prostate interstitial permanent brachytherapy with source assemblies are based upon the value of the seed air kerma strength, $S_K$. This is usually measured with well-type ionization chambers that, in most cases, are air-communicating chambers that require 
establishing the equilibrium with ambient conditions before measurements are done. The signal produced by these chambers is proportional to the air density and, therefore, measured values, taken at given pressure $P$ and temperature $T$, must be corrected to the standard atmospheric conditions using the well-known density correction factor:

\begin{equation}\label{ktp}
g_{0}\,=\, \frac{T}{T_0}\, \frac{P_0}{P}\, =\, \left( \frac{\rho}{\rho_0} \right)^{-1}\, .
\end{equation}

\noindent
Here $T_0=295.15\,$K, in the U.S.A., or $293.15\,$K, in Europe, and $P_0=1013.25\,$hPa ($760\,$mmHg) are the standard temperature and pressure. Note that the factor $g_0$ is the inverse of the ratio between the air density under actual, $\rho$, and standard, $\rho_0$, conditions. The density correction factor in equation (\ref{ktp}) applies if the Bragg-Gray theory holds for the chamber cavity. In these circumstances, the energy deposited by the secondary electrons produced in the chamber is proportional to the mass of air inside its active volume and their tracks are longer than the dimensions of that volume. Both the mass-stopping power of air for these secondary electrons and the electron tracks in the cavity are independent of the air density \cite{ICRU,Attix} and the corrected readout is given by:

\begin{equation}
\label{eq:MBG}
M^{\rm BG}_{\rm corr}\,=\,g_0(\rho) \,M_{\rm raw}(\rho)\, ,
\end{equation}

\noindent
where the ionization chamber response $M_{\rm raw}(\rho)$ represents the original (uncorrected) readout obtained from the ionization chamber at a given air density $\rho$.

However, some well-type ionization chambers, especially when used for assessing brachytherapy sources including low-energy radionuclides such as $^{103}$Pd, $^{125}$I or $^{131}$Cs, do not fulfill that condition and the corrected values turn out to be still density dependent. This behavior has been reported for several ionization chambers \cite{Griffin,Bohm,Russa,Russa2,Tornero} and an additional correction factor $g_1(\rho)$ has been included in order to make the final corrected readout density independent:

\begin{equation} \label{correction}
M_{\rm corr}\,=\, g_{1}(\rho)\, M^{\rm BG}_{\rm corr}(\rho)\, =\, g_{1}(\rho)\,g_0(\rho) \, M_{\rm raw}(\rho)\, ,
\end{equation}

Different functional dependences have been proposed for this additional factor. For example, 
Griffin {\it{et al.}} \cite{Griffin} found that $g_0$ over-corrected the result for low pressures in the case of the HDR 1000 Plus chamber (Standard Imaging, Middleton, USA) and, to account for the deviation, they proposed

\begin{equation} \label{Griff}
g^{\rm G}_{1}(P)\,=\,k_{1}\, P^{k_{2}} \, \equiv \, \left( \frac{P}{P_0} \right)^{k_2}\, ,
\end{equation}

\noindent
which guarantees that $g^{\rm G}_{1}(P_0)=1$. Here the parameter $k_2$ is specific of the particular seed and ionization chamber considered.

As a second example we mention here the SourceCheck type 34051 ionization chamber (PTW, Freiburg, Germany) which was analyzed by Tornero {\it et al.} \cite{Tornero}. This is a plane parallel ionization chamber widely used in Europe to determine $S_{K}$ in low dose rate brachytherapy. In this case, and contrary to the HDR 1000 Plus chamber, the factor $g_0$ produced an under-correction of the measurements for densities below $\rho_{0}$ and the additional correction factor 

\begin{equation}\label{1}
g^{\rm T}_{1}(\rho)\, = \,\left[A\,\left(\frac{\rho}{\rho_{0}} -1\right)\,+\,1\right]^{-1}
\end{equation}

\noindent
was introduced to eliminate the residual density dependence. The new factor $g^{\rm T}_{1}$ is inversely proportional to the density and the parameter $A$, which was determined by a linear regression of the obtained measurements, appeared to be dependent on the particular specimen of SourceCheck ionization chamber used.

Recently, PTW designed the SourceCheck 4pi type 33005, a new version of the SourceCheck 34051, which has been discontinued. The purpose of the present work is to study the air density dependence of the response of this new ionization chamber. Three SourceCheck 4pi chambers, provided to us by the manufacturer, and three HDR 1000 Plus chambers were used in the same measuring conditions and this permitted to carry out a complete comparison among the two chamber models and the specimens of the same model.

\section*{Material and methods}

\subsection*{Instrumentation}

According to the manufacturer, the new PTW SourceCheck 4pi 33005 is a well-type ionization chamber with Al walls and an Al-coated polystyrene round pipe inner electrode. Its sensitive volume is $116\,$cm$^3$, it has a length of $15.00\pm 0.02\,$cm and a diameter of $3.205 \pm 0.005\,$cm. Three different chambers of this model were provided by the manufacturer to perform the present study. This permitted us to analyze the possible specimen dependence for this new ionization chamber model. Measurements were obtained by using a PTW UNIDOS electrometer.

We also used three HDR 1000 Plus ionization chambers for studying the variability of their responses with pressure and comparing the results obtained with those found for the SourceCheck 4pi chamber and those quoted in previous works \cite{Griffin}. The HDR 1000 Plus is an open well chamber constructed with an inner wall made of butyrate and lined with a thin aluminum foil, an aluminum collecting electrode and an aluminum outer wall. It is $16\,$cm high, with a diameter of $10\,$cm. The manufacturer indicates that its active volume is $275\,$cm$^3$. These chambers were used with a MAX4000 electrometer, manufactured by Standard Imaging. 

Sources including two different radionuclides were used. The first one, $^{125}$I, decays by electron capture and the average energy of the emitted photons is about 28 keV. The second one, $^{90}$Sr, is a high energy beta emitter that decays to $^{90}$Y, another radionuclide which undergoes beta decay to the stable isotope $^{90}$Zr. Sr-Y sources are commonly used to check the stability of ionometric equipment and in the present work has been used to check if the overcorrection disappears when measuring a high-energy source, for which Bragg-Gray theory is applicable to the ionization chamber. A $^{125}$I selectSeed$^{\rm TM}$ seed (Isotron Isotopentechnik GmbH, Solingen, Germany), with a nominal air kerma strength value $S_{K}=0.610\,$cGy$\,$cm$^{2}\,$h$^{-1}$, and a $^{90}$Sr PTW type 8921 source, with an activity of $18.3\,$MBq, were used in our experiment. 

\subsection*{Experiments}

The air density dependence of the ionization chamber response was studied by changing the pressure of the air surrounding the ionization chamber. The well chamber was introduced in a homemade pressure chamber, built up in PMMA and with dimensions of $30\times 24\times 24\,$cm$^{3}$. A groove permitted the connection of the ionization chamber studied and the electrometer situated outside the pressure chamber. Pressure changes were produced by means of a vacuum pump connected to the pressure chamber through a suction nozzle inserted into one of the walls. A thermometer-hygrometer-barometer THB~40 (PCE Instruments UK Ltd, Southampton, UK) allowed measuring pressure and temperature inside the pressure chamber and also the relative humidity. The same setup was used in \cite{Tornero} to study the behavior of the old SourceCheck 34051 ionization chamber.

Measurements for pressures varying from $746.6$ to $986.6\,$hPa ($560-740\,$mmHg), in steps of $\sim 27\,$hPa ($20\,$mmHg) were carried out. Three readouts were taken for each pressure value. Each readout took about 1 minute and was obtained after reaching the equilibrium between the air pressure inside and outside the ionization chamber. As it occurs for Farmer-type chambers \cite{Robinson}, the equilibrium was accomplished in a few seconds and, before each measurement, we waited around $10-15\,$s after the reading of the barometer was stabilized. The calibration of the barometer indicated a systematic deviation of  $0.4\,$hPa ($=0.3\,$mmHg) that was added to the values considered in the experiment to carry out the corresponding corrections. 

Though it has been shown that reaching thermal equilibrium in air may require longer times, of the order of several minutes for Farmer-type chambers  \cite{Kubo}, in our case, the measuring setup was placed in the laboratory several hours before the measurements were carried out. In addition, temperature remained almost constant during the whole experiment, which took less than two hours, with a maximum variation of $0.6\,^{\circ}$C. This permitted to reach the equilibrium with the environment in times much shorter than those indicated in the ESTRO booklet no. 8 \cite{ESTRO}. On the other hand, the calibration of the thermometer indicated a zero-offset of $0.3\,^{\circ}$C that was subtracted from the temperature values in each measurement.

Finally, relative humidity remained almost constant during the whole experiment, varying between $42\%$ and $48\%$.

\subsection*{Data analysis}

First, the raw readouts were corrected with $g_{0}$ to obtain the $M^{\rm BG}_{\rm corr}(\rho)$ values (see equation (\ref{eq:MBG})). To obtain the additional correction factor $g_1(\rho)$, the function 
\begin{equation}
y(\rho)\, =\, \frac{M_0}{g_1(\rho)}
\end{equation}
was fitted to the $M^{\rm BG}_{\rm corr}(\rho)$ data obtained for each ionization chamber and source considered. Here
$M_0$ is the reference value under normal conditions of temperature and pressure ($293.15\,$K and $1013.25\,$hPa). Two different approaches for $g_{1}$ were tested. The first one was that introduced by Tornero {\it{et al.}} \cite{Tornero}, which is given in equation (\ref{1}). The second one was that proposed in the PTW user manual \cite{PTW}, which is derived from the work of Griffin {\it et al.} \cite{Griffin} and which is defined as a potential function:

\begin{equation}\label{fitptw}
g_1^{\rm PTW}(\rho)\, =\, \left(\frac{\rho}{\rho_{0}}\right)^{u} \, .
\end{equation}

Fits were carried out by minimizing the corresponding value of $\chi^2$ in the chi-square test using the Levenberg-Marquardt method \cite{Press}. The fitting parameters were $M_0$ and $A$ in the first case and $M_0$ and $u$ in the second one. Comparisons between
\begin{equation}
\bar{M}(\rho)\,=\,\frac{M^{\rm BG}_{\rm corr}(\rho)}{M_{0}} \, = \, g_{0}(\rho)\, \frac{M_{\rm raw}(\rho)}{M_{0}} \, ,
\label{eq:Mbar}
\end{equation}
and $[g_1(\rho)]^{-1}$ were done to analyze the capabilities of the two functional forms considered for $g_1$.

\subsection*{Uncertainties}

Table \ref{tabla:tabla2} shows the relative uncertainties affecting the experiments performed in the present study with the pressure chamber. Total uncertainties in $\bar{M}$ have been computed as the quadratic sum of type A and type B uncertainties. The first ones come from the variability of the data and are calculated as the coefficient of variation of the three measurements taken for each air density value.

\begin{table}[htb]
\begin{center}
\begin{tabular}{ccccc}
\hline \hline
 &~~&  \multicolumn{3}{c}{type B} \\ \cline{3-5}
   type A  && thermometer & barometer & electrometer \\ \hline
  $0.4\%\,^{\rm a}$& & $0.1\,^{\rm o}{\rm C}\, ^{\rm b}$ & $0.78\,$hPa$\,^{\rm c}$ & $0.01\%\,^{\rm b}$ \\
\hline \hline
\end{tabular}
\caption{\small Relative uncertainties (with a coverage factor $k=1$) affecting the experimental measurements carried out with the pressure chamber in the present work. $^{\rm a}$Average coefficient of variation of raw measurements. $^{\rm b}$Manufacturer specifications. $^{\rm c}$Quadratic sum of the uncertainty indicated in manufacturer specifications and the pressure variation observed in the pressure chamber.}
\label{tabla:tabla2}
\end{center}
\end{table}

Type B uncertainties in our experiment came from those linked to the electrometer, the barometer and the thermometer.
The thermometer and electrometer uncertainties were those quoted in the specifications of the respective manufacturers. For the experiments performed with the pressure chamber, two components were considered for the barometer type B uncertainty. Throughout the experiment, the pressure could change within $\pm 1\,$mmHg. Assuming a uniform probability distribution of pressures, the associated uncertainty (with a coverage factor $k=1$) is given by \cite{WG1}:

\begin{equation}
u(P)\,=\, \sqrt{\frac{(2\,{\rm mmHg})^{2}}{12}}\,=\, \frac{1}{\sqrt{3}}\,{\rm mmHg}\, \approx \, 0.77\,{\rm hPa}\, .
\end{equation}
This was added in quadrature with the standard uncertainty quoted in the manufacturer's specification, giving a total type B uncertainty of $0.78\,$hPa $(k = 1)$ as shown in table \ref{tabla:tabla2}.

\section*{Results and discussion}

%\subsection{Pressure variations}

\begin{figure}[b!]
\centering
\includegraphics[width=0.4\textwidth]{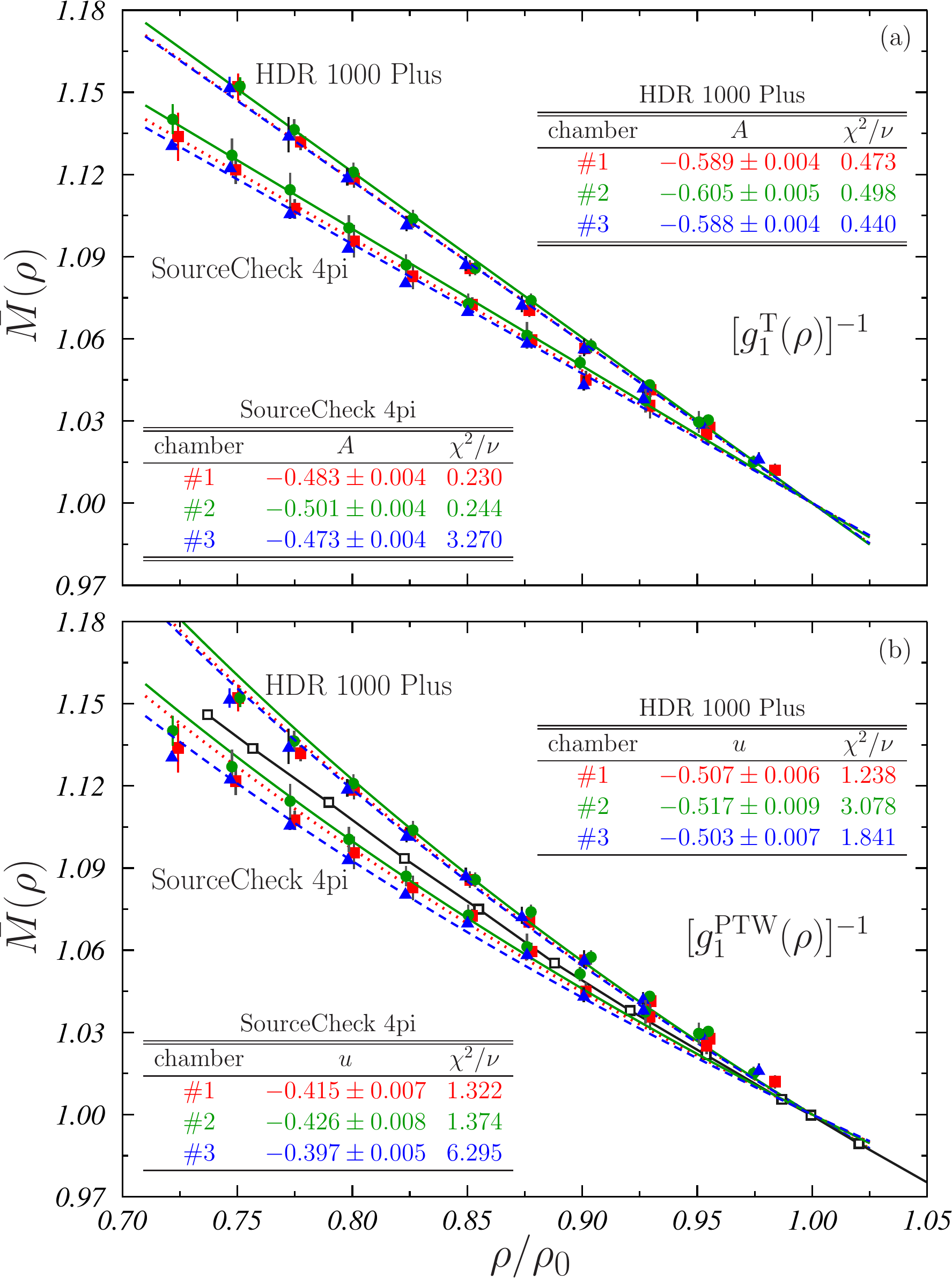}
\vspace*{-0.5cm}
\caption{\small Values of $\bar{M}(\rho)$ (see equation (\ref{eq:Mbar})), as a function of $\rho/\rho_0$, for the HDR 1000 Plus and the SourceCheck 4pi ionization chambers, for $^{125}$I selectSeed$^{\rm TM}$ seeds. The experimental data for the three specific chambers considered in each case (red squares for chamber \#1, green circles for chamber \#2 and blue triangles for chamber \#3) are plotted together with the corresponding linear regressions (figure (a)) done by using $[g_1^{\rm T}(\rho)]^{-1}$, defined in equation (\ref{1}), and the fits (figure (b)) performed with $[g_1^{\rm PTW}(\rho)]^{-1}$, given in equation (\ref{fitptw}), (red dotted lines for chamber \#1, green solid lines for chamber \#2 and blue dashed lines for chamber \#3). The fitting parameter $A$ and $u$, as well as the $\chi^2$ values per degree of freedom are shown in the inset tables for all fits. Uncertainties in the data points and the fitting parameter values are given with a coverage factor $k=1$. Open squares in figure (b) represent the data provided by Griffin {\it et al.} \cite{Griffin} for the HDR 1000 Plus chamber.} 
\label{fig:1}
\end{figure}

In figure \ref{fig:1}a, the results of the measurements for the $^{125}$I selectSeed$^{\rm TM}$ seeds made with the three HDR 1000 Plus and three SourceCheck 4pi chambers are shown. The corresponding linear fits using $[g_1^{\rm T}(\rho)]^{-1}$ (see equation (\ref{1})) are also plotted. 

The first point that deserves a comment is that related to the air density dependence of the response of the new SourceCheck 4pi chamber. Once the usual temperature and pressure correction is carried out, $\bar{M}(\rho)$ was still dependent on the air density, the ionization chamber response being overestimated at air densities $\rho < \rho_0$. This dependence is similar to that occurring for the HDR 1000 Plus chamber and opposite to what has been obtained in the case of the old SourceCheck for which an under-correction occurred after considering $g_0(\rho)$ \cite{Tornero}. The slopes for the HDR 1000 Plus chambers are larger (in absolute value) than for the SourceCheck 4pi chambers. To have a quantitative idea of the overestimation observed, it is worth noting that it reaches $\sim 4\%$ in cities situated at a high altitude as ours (700 m above mean sea level, AMSL).

In figure \ref{fig:1}b the same $\bar{M}(\rho)$ experimental data of figure \ref{fig:1}a are plotted together with the fits obtained by using $[g_1^{\rm PTW}(\rho)]^{-1}$, which is defined in equation (\ref{fitptw}). According to the results quoted in the inset tables, it is evident that these fits are worse than the linear fits provided by $[g_1^{\rm T}(\rho)]^{-1}$. As we can see, the values of $\chi^2$ per degree of freedom are significantly smaller in the latter case. In figure \ref{fig:1}b, also the values obtained by Griffin {\it et al.} \cite{Griffin} for the HDR 1000 Plus chamber are shown with open squares. They show a systematic deviation with respect to our results.

As shown in figure \ref{fig:1}a, the values measured for the three SourceCheck 4pi 33005 specimens used are almost overlapping, independently of the value of $\rho/\rho_0$. The behavior is very similar to that found for the HDR 1000 Plus chamber, though the spread is slightly larger. Nevertheless, it is worth pointing out that these results are again contrary to those found for the old SourceCheck 34051 chamber, which were strongly dependent on the particular specimen used to perform the measurements \cite{Tornero}.

\begin{figure}[h!]
\centering
\includegraphics[width=0.4\textwidth]{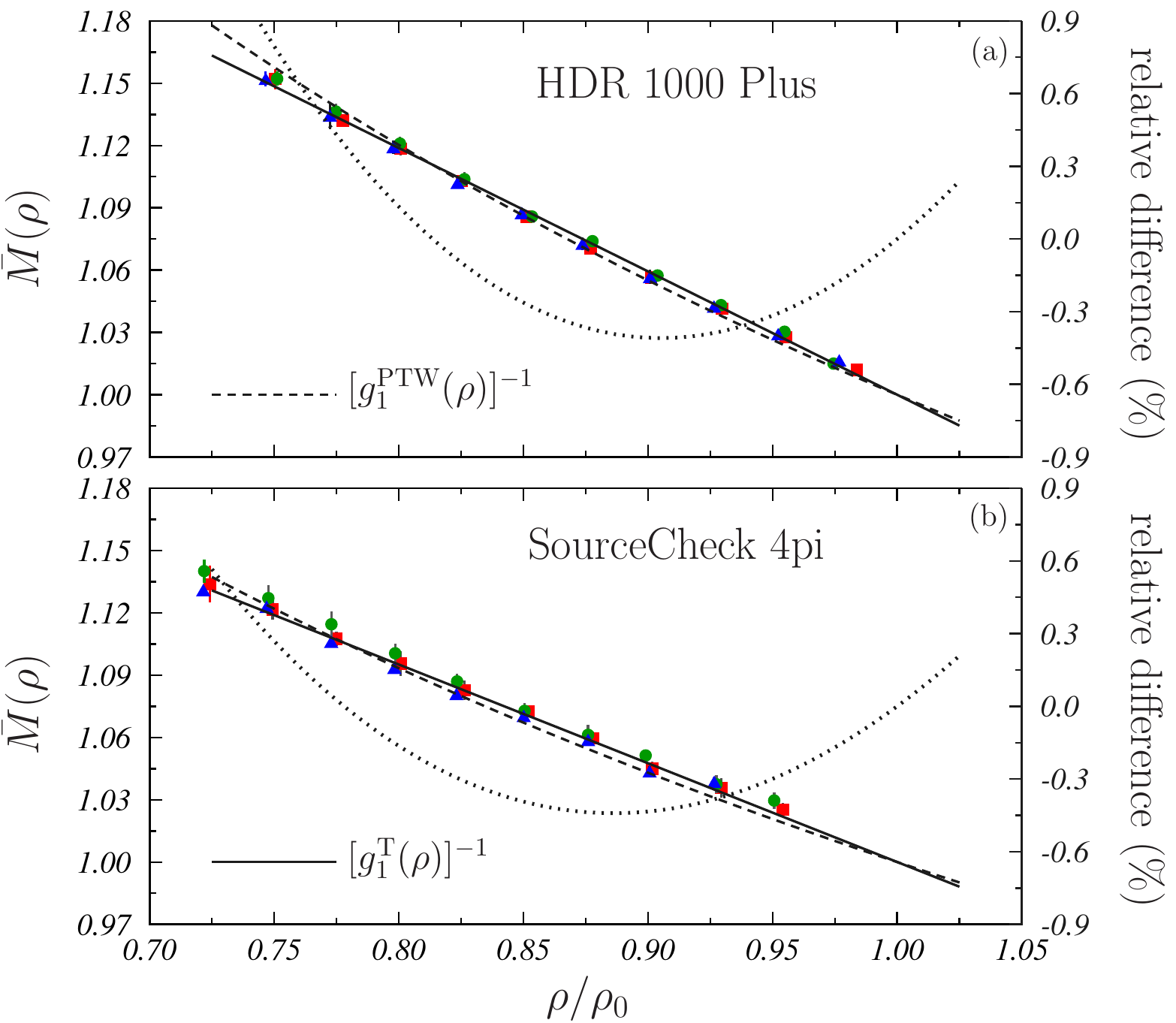}
\caption{\small Values of $\bar{M}(\rho)$ (see equation (\ref{eq:Mbar})), as a function of $\rho/\rho_0$, for the HDR 1000 Plus (figure (a)) and the SourceCheck 4pi (figure (b)) ionization chambers, when $^{125}$I selectSeed$^{\rm TM}$ seeds are measured. The fits to all experimental data available, done by using $[g_1^{\rm PTW}(\rho)]^{-1}$ (dashed line) and $[g_1^{\rm T}(\rho)]^{-1}$ (solid line), are shown. Dotted curves represent the relative differences between both fits (right ordinate axis). Uncertainties in the data points are given with a coverage factor $k=1$.} 
\label{fig:4}
\end{figure}

Taking this into account, we fitted the data again but this time we included all experimental data available for each chamber model. These global fits are shown in figure \ref{fig:4} for the HDR 1000 Plus (figure (a)) and the SourceCheck 4pi (figure (b)) ionization chambers. The dashed curves represent the results obtained by using $[g_1^{\rm PTW}(\rho)]^{-1}$ and the solid lines those found with $[g_1^{\rm T}(\rho)]^{-1}$. Dotted curves represent the relative differences between both fits and are referred to the right ordinate axes. The fitting parameters as well as the $\chi^2$ values per degree of freedom are given in table \ref{tabla:tabla}.

\begin{table}[!b]
\begin{center}
\begin{tabular}{ccccccccc}
\hline \hline
&&&&\multicolumn{3}{c}{SourceCheck 4pi} && HDR 1000 Plus \\ \cline{5-7} \cline{9-9}
\multicolumn{3}{c}{fit} & ~~ & $^{125}$I selectSeed$^{\rm TM}$ & ~~& $^{90}$Sr PTW 8921 &~~& $^{125}$I selectSeed$^{\rm TM}$ \\\hline %\cline{5-5} \cline{7-7}
 $[g_1^{\rm T}(\rho)]^{-1}$ && A                         && $-0.476\pm 0.003$ && $-0.017 \pm 0.001$ && $-0.594\pm 0.003$  \\
                             && $\chi^2/\nu$        && 1.471                      && 0.503                       && 0.583                       \\\hline
 $[g_1^{\rm PTW}(\rho)]^{-1}$ && $u$                     && $-0.400\pm 0.003$ && $-0.015\pm 0.001$  && $-0.509\pm 0.004$ \\
                             && $\chi^2/\nu$        && 3.484                      && 0.497                       && 2.047                       \\\cline{3-9}
                           && $u$  && $-0.401$ \cite{PTW}                  && -0.019 \cite{PTW}                        && $-0.545$  \cite{Griffin}    \\\hline\hline
\end{tabular}
\caption{\small Values of the parameters describing the dependence of the chamber response on air density for the two different types of well chamber used in this study. The results quoted correspond to the fits performed by including all data available for each chamber model. Results for the $^{125}$I and $^{90}$Sr sources used are shown. Uncertainties have a coverage factor of $k=1$. The $u$-values (see equation (\ref{fitptw})) quoted in the PTW manual \cite{PTW} for both sources are given in the case of the SourceCheck 4pi. For the HDR 1000 Plus, the value $u=1-k_2$, obtained from the $k_2$ found by Griffin {\it et al.} \cite{Griffin} for the $^{125}$I Amersham 6711 seed, is shown. The dimensions and materials of this $^{125}$I source are similar but not identical to those of the  $^{125}$I selectSeed$^{\rm TM}$.}
\label{tabla:tabla}
\end{center}
\end{table}

These results show again that the fits found by using $[g_1^{\rm T}(\rho)]^{-1}$ are better than those obtained for $[g_1^{\rm PTW}(\rho)]^{-1}$.
In fact, the global fits including all the $^{125}$I data available show, in the latter case, $\chi^2 /\nu$ values of 2.047 and 3.484 for the HDR 1000 Plus and the SourceCheck 4pi chambers, respectively, while these values reduce to 0.583 and 1.471 in the case of the linear fit. In the range of air densities analyzed, the relative differences between both fits are within $\pm 0.6\%$ except for the HDR 1000 Plus chamber at values of $\rho/\rho_0$ below $\sim 0.76$. 

It is worth pointing out that the value of the parameter $u$ recommended by PTW \cite{PTW} is in agreement with our result within the uncertainties of our calculation. On the other hand, the value of the fitting parameter $u$ quoted by Griffin {\it et al.} \cite{Griffin} for the HDR 1000 Plus chamber differs from ours by 7\%. However, the differences between the corresponding $\bar{M}(\rho)$ values become larger than 1\% for relative air densities smaller than about 0.78 (see figure \ref{fig:1}b). These differences do not change if the zero-offsets we found for pressure and temperature are ignored.

The reasons for the the overcorrection observed in figure \ref{fig:1}, after including the factor $g_0(\rho)$, have been widely studied and different effects have been identified to contribute \cite{Bohm,Russa}. The main one is linked to the range of the secondary electrons generated in the chamber: the lower the energy of the photons emitted by the source, the smaller the maximum energy and the range of these secondary electrons. When that range is of the order of the characteristic length of the active volume of the ionization chamber, which is the case for the $^{125}$I sources, a large number of the electrons produced stops in the chamber active volume. Changes in the air density do not produce a modification of the total energy deposited, which remains almost constant. In these circumstances, the factor $g_0$ produces an overcorrection of the readout at low air densities and vice versa.

However, a second effect, opposite to this one, must be taken into account. If the air density increases, more electrons will stop inside the cavity and  the sharp enhancement of the stopping power for electrons with kinetic energies close to zero will produce an acute rise of the energy deposited that $g_0$ cannot describe adequately: the readout will be overestimated (underestimated) for air densities above (below) $\rho_0$.

The dominance of one of the two mechanisms is due to a delicate balance of various components: the spectra of the secondary electrons produced in the ionization chamber, the dimensions of its active volume, the variations in the cross-sections of the photoproduction of electrons in the chamber construction materials, etc. The geometrical differences between the new and old models of the SourceCheck ionization chamber may be considered as responsible of the strongly unpaired responses shown by these two chambers; on the other hand, the similar responses obtained for the SourceCheck 4pi and the HDR 1000 Plus can be ascribed to their similar design.

\begin{figure}[!t]
\centering
\includegraphics[width=0.45\textwidth]{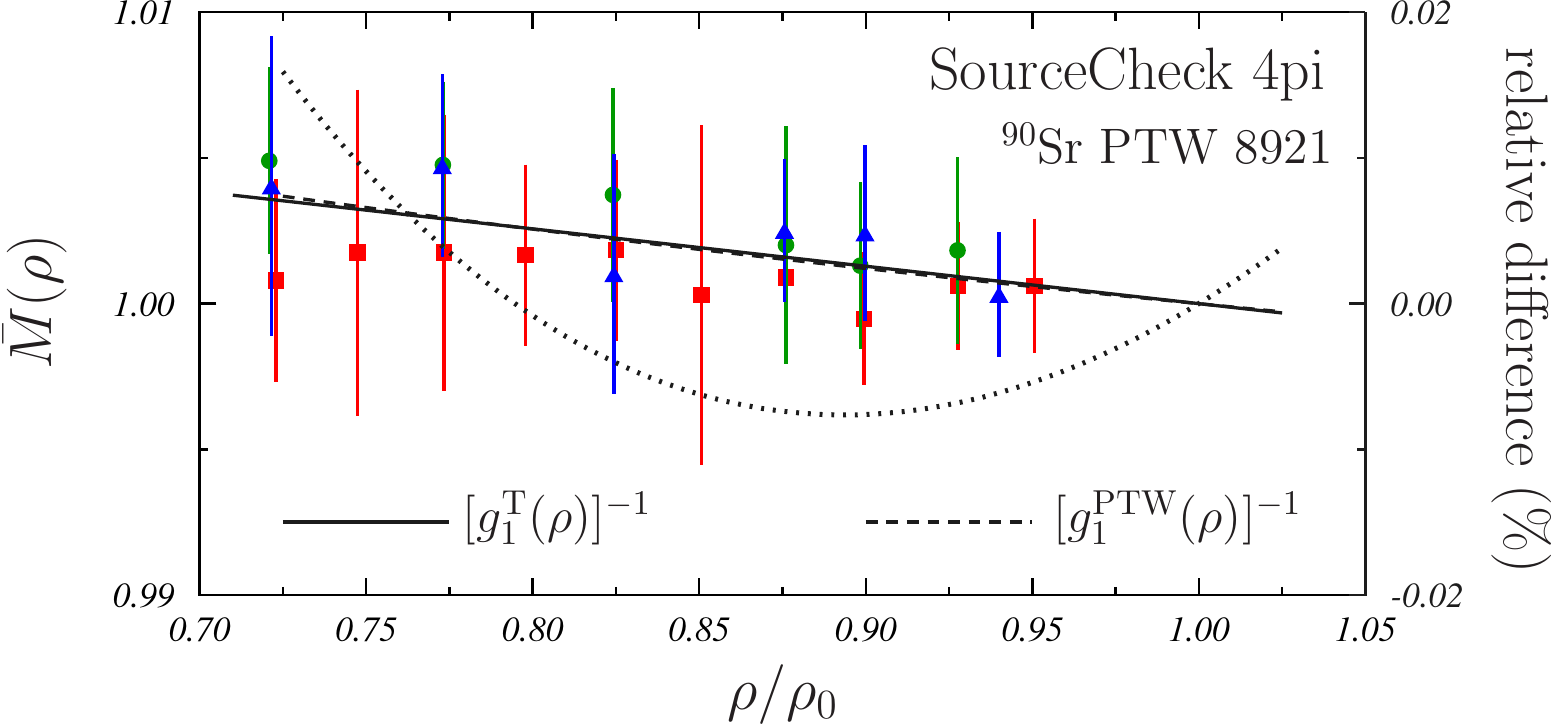}
\caption{\small Values of $\bar{M}(\rho)$ (see equation (\ref{eq:Mbar})), as a function of $\rho/\rho_0$, for the SourceCheck 4pi ionization chamber, when $^{90}$Sr PTW 8921 sources are measured. The fits to all experimental data available, done by using $[g_1^{\rm PTW}(\rho)]^{-1}$ (dashed line) and $[g_1^{\rm T}(\rho)]^{-1}$ (solid line), are shown. The dotted curve represents the relative difference between both fits (right ordinate axis). Uncertainties in the data points are given with a coverage factor $k=1$.} 
\label{fig:2}
\end{figure}

To complete the present analysis, we have performed a series of measurements with the three SourceCheck 4pi chambers available for the $^{90}$Sr PTW 8921 sources. The corresponding $\bar{M}(\rho)$ values are shown in figure \ref{fig:2} together with the global fits (including all data) performed with the two approaches we are considering in the present work. The first point to be noted is that the dependence of the chamber response on air density is strongly reduced, confirming our expectation according to the higher energies of the betas emitted after the decay of the $^{90}$Sr radionuclide. On the other hand, both fits provide very similar results and the relative differences are very small, within $\pm 0.02\%$. The value of the fitting parameters and the $\chi^2$ values per degree of freedom are shown in table \ref{tabla:tabla}. In this case, the $u$-value recommended by PTW \cite{PTW} is not statistically compatible with our result.

The importance of an accurate characterization of the dependence on air density of the response of this kind of ionization chambers can be understood even better if one takes into account the uncertainties that can be accepted in the determination of the air kerma strength for low dose rate brachytherapy sources. According to AAPM \cite{DeWerd}, the uncertainties affecting the various measurements performed with well chambers to transfer the air kerma strength standard to clinic result in a minimum uncertainty of $2.56\%$ (with a coverage factor $k=2$). For chambers whose response shows a dependence with air density, neglecting it or performing an incorrect characterization of its effect counts double: in the calibration process itself, performed by the corresponding accredited dosimetry calibration laboratory, and in the clinical measurement performed by the end user. Assuming, for example, the usual atmospheric conditions in our city ($700\,$m AMSL), using the SourceCheck 4pi without the additional air density correction would overestimate the air kerma strength by $\sim 4\%$. This value is of the order of the maximum difference accepted between the air kerma strength indicated by the source manufacturer and that verified by the end user. According to TG138 \cite{DeWerd}, the source acceptance criterion is $3.4\%$. This reinforces the necessity of an adequate knowledge of the correction factors of any ionization chamber used in the clinical practice.

\section*{Conclusions}

In this work we have studied the dependence on the air density of the response of the new SourceCheck 4pi 33005 ionization chamber manufactured by PTW. We have found that it shows a residual air density dependence when measuring low-energy photon radiation. This dependence follows the same trend as that observed for the Standard Imaging HDR 1000 Plus ionization chamber of Standard Imaging and an inverse trend to the behavior shown by the old PTW SourceCheck 34051 chamber. The read-outs are overcorrected at densities lower than that at standard atmospheric conditions. This additional dependence can be described as a linear function of the air density, and this kind of fit describes the data much better than other models which are based on potential functions. The fitting function is universal and does not depend on the particular ionization chamber used, as the old SourceCheck chamber. 

\section*{Acknowledgments}

We would like to acknowledge PTW for providing us with the new SourceCheck 4pi ionization chambers and
Salvador Garc\'{\i}a Pareja and Luis I. Ramos Garc\'{\i}a for allowing us to use their HDR 1000 Plus ionization chambers. This work has been partially supported by the Spanish Ministerio de Ciencia y Competitividad (FPA2015-67694-P), European Regional Development Fund (ERDF) and by the Junta de Andaluc\'{\i}a (FQM0387).

%\section*{References}

\end{document}